\documentclass[12pt]{iopart}

\usepackage[colorlinks,linkcolor=blue,citecolor=blue,urlcolor=blue]{hyperref}
\usepackage{eqnarray}
\usepackage{graphicx}
\usepackage{upgreek}

\renewcommand{\vec}[1]{\mathbf{#1}}

\newcommand{\dd}{\mathrm{d}}

\newcommand{\iu}{\mathrm{i}}

\newcommand{\hc}{\hat{c}}
\newcommand{\eqref}[1]{(\ref{#1})}

\begin{document}

\title{Femtosecond dynamics of correlated many-body states in
  C$_{60}$ fullerenes}

\author{Sergey Usenko$^1$, Michael Sch\"{u}ler$^2$, Armin
  Azima$^{3,4,5}$, Markus Jakob$^{1,5}$, Leslie L Lazzarino$^3$,
  Yaroslav Pavlyukh$^2$, Andreas Przystawik$^1$, Markus
  Drescher$^{3,4,5}$, Tim Laarmann$^{1,5}$ and Jamal Berakdar$^2$}
\address{$^1$ Deutsches Elektronen-Synchrotron DESY, Notkestr.~85,
  22607 Hamburg, Germany} \address{$^2$ Institut f\"{u}r Physik,
  Martin-Luther-Universit\"{a}t Halle-Wittenberg, 06099 Halle,
  Germany} \address{$^3$ Department of Physics, University of Hamburg,
  22761 Hamburg, Germany} \address{$^4$ Center for Free-Electron Laser
  Science CFEL, DESY, 22607 Hamburg, Germany} \address{$^5$ The
  Hamburg Centre for Ultrafast Imaging CUI, Luruper Chaussee~149,
  22761 Hamburg, Germany} \ead{tim.laarmann@desy.de}
\ead{jamal.berakdar@physik.uni-halle.de}

\begin{abstract}
  Fullerene complexes may play a key role in the design of future
  molecular electronics and nanostructured devices with potential
  applications in light harvesting using organic solar cells. Charge
  and energy flow in these systems is mediated by many-body
  effects. We studied the structure and dynamics of laser-induced
  multi-electron excitations in isolated C$_{60}$ by two-photon
  photoionization as a function of excitation wavelength using a
  tunable fs UV laser and developed a corresponding theoretical
  framework on the basis of \emph{ab initio} calculations. The
  measured resonance line width gives direct information on the
  excited state lifetime. From the spectral deconvolution we derive a
  lower limit for purely electronic relaxation on the order of
  $\tau_\mathrm{el}=10^{+5}_{-3}$\,fs. Energy dissipation towards
  nuclear degrees of freedom is studied in time-resolved
 techniques. The evaluation of the non-linear autocorrelation trace
  gives a characteristic time constant of \mbox{$\tau_\mathrm{vib}=400\pm100$\,fs} for the exponential decay. In line with the experiment, the
  observed transient dynamics is explained theoretically by nonadiabatic
  (vibronic) couplings involving the correlated electronic, the nuclear degrees of freedom (accounting for the
  Herzberg-Teller coupling), and their interplay.
\end{abstract}

\maketitle

\section{Introduction}
Molecular junctions, molecular
transistors and organic solar cells rely on  charge transport channels
with negligible energy dissipation during the  carriers propagation
time. In nanostructured materials and molecular complexes the
characteristic timescale is determined by the long-range polarization
interaction and by the formation and breaking of chemical bonds
mediated by the electronic and nuclear motion. Transient structures and
dynamics on the femto and sub-femtosecond timescale  is the focus of
ultrafast spectroscopy. Time-resolved experiments using femtosecond
(fs) laser pulses unravel the dynamic response of promising materials
that could serve for instance as molecular building blocks for organic
photovoltaics. Polymer solar cells are commonly composed of a
photoactive film of a conjugated polymer donor and a fullerene
derivative acceptor\,\cite{sariciftci_photoinduced_1992,yu_polymer_1995,scholes_excitons_2006},
which makes use of the fullerenes' unique ability to form stable
C$_{60}^-$ anions. Electron correlation plays an important role in the
formation of four bound states of the fullerene anion\,\cite{klaiman_extreme_2013,klaiman_all_2014,voora_nonvalence_2014}.
In fact, electronic correlations are responsible for the binding of the 2A$_\mathrm{g}$ state, whereas the bindings of the states
2T$_\mathrm{1u}$, 2T$_\mathrm{2u}$ and 2T$_\mathrm{1g}$ are less affected by electronic correlations (cf.\ Ref.\,\cite{klaiman_extreme_2013} and further references therein).

With its special structure consisting of 174 nuclear degrees of
freedom, 60 essentially equivalent delocalized $\pi$ electrons, and
180 structure-defining localized $\sigma$ electrons, neutral C$_{60}$
serves as a model for a large -- but still finite -- molecular system
with many electronic and nuclear degrees of freedom. Because of the
large charge conjugation, its finite ''energy gap'', and quantum
confinement of electronic states, C$_{60}$ may be viewed as an
interesting intermediate case between a molecule and a condensed
matter system. In fact, applying solid-state concepts to the valence
''Bloch electrons'' on the C$_{60}$ sphere results in an ''angular
band structure''\,\cite{pavlyukh_angular_2009} from which other
relevant quantities (such as plasma frequencies and group velocities)
can be extracted. Photophysical studies of fullerenes using fs laser
fields cover the whole range from atomic through molecular to solid
state physics\,\cite{hertel_ultrafast_2005,jacquemin_direct_1998,link_electron_2000}.
The molecular response is truly a multi-scale phenomenon. It ranges
from attosecond dynamics in electronic excitation and ionization to
statistical physics describing thermalization processes. So,
light-induced processes in fullerenes cover more than 15 orders of
magnitude in time\,\cite{lepine_multiscale_2015}.

Using low-temperature scanning tunneling microscopy (LT-STM) of
C$_{60}$ molecules deposited on copper surfaces Feng \emph{et al.}
observed tunneling through electronic states that possess nearly
atom-like character\,\cite{feng_atomlike_2008}.  These ''superatom''
molecular orbitals (SAMOs), also discussed below, have a well-defined
symmetry and can be characterized by the nodal structure (principle
quantum numbers) $n$ and angular momentum quantum numbers $L$\,\cite{feng_electronic_2011}. In addition, these virtual states show a
remarkable stability\,\cite{pavlyukh_communication:_2011}, i.e., their
initial stage of decay proceeds substantially slower than other states
(even LUMO or HOMO) which qualifies them as robust channel for hot
electron transport. From a molecular physics point of view, SAMOs can
be regarded as low-lying, mixed valence Rydberg states\,\cite{reisler_interacting_2009} that exhibit substantial electron
density inside the hollow sphere. They form chemical bonds affected by
hybridization when the system is excited optically or probed in STM in
deposited nanostructures. The resulting free-electron bands in
self-assembled one-dimensional wires or two-dimensional quantum wells
are holding great promise for unique applications in molecular
electronics\,\cite{zhao_non-nuclear_2014}, but also for new
functionalities such as current carrying states and hence
nanometer-sized magnetic field generators\,\cite{watzel_optical_2016}. We note in passing that recently SAMO
states have also been observed in planar, non-fullerene materials\,\cite{zoppi_buckybowl_2015,dougherty_band_2012}, as well as in
isolated C$_{60}$ fullerenes\,\cite{johansson_angular-resolved_2012,mignolet_probing_2013,bohl_relative_2015},
where they are energetically located below the known high-lying
Rydberg states\,\cite{boyle_excitation_2001,boyle_excitation_2005}.

It is of fundamental importance for designing fullerene derivatives as
building blocks for solid state chemistry to go beyond the
characterization of \emph{static} many-body electronic structure\,\cite{roy_nanoscale_2013}. In particular for optimization and control
of the charge flow and energy dissipation, rigorous \emph{dynamic} studies on
the fs time scale using ultrafast lasers are indispensable but, still
in their infancy\,\cite{shchatsinin_ultrafast_2008,li_coherent_2015}.
Additionally, time-dependent density-functional theory (TDDFT)
calculations of the absorption and photoelectron spectra, accounting
for full structural analysis, were performed\,\cite{schuler_disentangling_2015,wopperer_progress_2015} and
constitute the basis for the further development presented below.

Accessing energy dissipation upon the excitation of correlated
many-body states in gas phase C$_{60}$ became feasible recently which
allows to connect the coherent quantum\,\cite{schuler_disentangling_2015,schuler_electron_2016}, classical and
statistical mechanisms\,\cite{garcia_de_abajo_momentum_2004}. It is
known from experiments with optical lasers that electron
thermalization mediated by inelastic electron--electron collisions
takes place on a time scale of $50{-}100$\,fs (see\ \cite{lepine_multiscale_2015} and references therein). This is where
the present experimental and theoretical work comes into play. Here,
the objective is to reduce the complexity of laser-induced
multi-photon processes by populating highly-excited many-body states
in a resonant one-photon transition in the ultraviolet (UV) spectral range
at low laser peak intensity on the order of $3.5\times10^{10}$\,W/cm$^2$. This allows for a rather detailed
probing of the correlated electron dynamics in highly excited
states. The study is based on a resonance-enhanced multi-photon
ionization (REMPI) scheme, i.e., two-photon photoemission (2PPE)
spectroscopy as depicted in Fig.\,\ref{fig1}(a). The photoionization
yield recorded for resonant excitation is enhanced as compared to an
experiment performed in the off-resonance regime. Thereby, we trace
the time-dependent electronic structure of intermediate states free
from any perturbation caused by metallic substrates affecting the
energetics in LT-STM experiments. Furthermore, REMPI on gas phase
fullerenes provides information on the neutral molecule whereas 2PPE
and LT-STM essentially probe the binding energy and density of states
of an anion deposited on a metal surface. Our experiments are compared
to \emph{ab initio} calculations.

This paper is organized as follows. In Sec.\,\ref{sec:optabs}
many-body states below the C$_{60}$ ionization threshold are
calculated as guideline for the two-photon photoemission
experiments. Sec.\,\ref{sec:exp} describes some experimental details
with a focus on the tunable fs laser system in the UV spectral range used for the time-resolved studies. Excitation energy
dependent mass spectra are evaluated in Sec.\,\ref{sec:2ppe} and
discussed in terms of resonance-enhanced ionization and excited state
lifetimes. Sec.\,\ref{sec:relax} concentrates on the time-resolved
experiments. A detailed theoretical analysis of the experimental data
is given in Sec.\,\ref{sec:T2ppe} followed by a short summary and outlook. In \ref{app:elvib} we provide details on how the electron-vibron coupling matrix elements were computed and the master equation in Lindblad form is derived in \ref{app:lindblad}.

\section{Optical excitations \label{sec:optabs}}
The first step of a REMPI or 2PPE experiment entails the calculation
of excited states, which are typically more difficult to describe than
ground-state properties. Often exact diagonalization (full
configuration interaction) is not feasible for large systems in which
case one may resort to only a few methods: TDDFT\,\cite{marques_fundamentals_2012}, equation-of-motion (EOM) quantum
chemistry methods\,\cite{bartlett_coupled-cluster_2007}, and many-body
perturbation theory (MBPT) based on a Green's function formulation\,\cite{aryasetiawan_gw_1998}. For its reduced computational cost as
compared to the other methods, we have employed the linear-response
TDDFT approach (Casida's method)\,\cite{chong_recent_1995}.

As a first step we calculated the Kohn-Sham (KS) orbitals using the
{\sc octopus} package\,\cite{andrade_real-space_2015}. A modified
version of the asymptotically corrected functional by Leeuwen and
Baerends\,\cite{van_leeuwen_exchange-correlation_1994}, which was shown
to considerably improve excited-state
properties\,\cite{schipper_molecular_2000}. In order to account for a
multitude of highly-excited states, we have chosen a relatively large
box to which all KS states $\phi_{i}(\vec r)$ are confined (a sphere
with radius 12\,{\AA} with uniform grid spacing of 0.15\,{\AA}). This
ensures that higher virtual orbitals (including the SAMOs) are well
represented. After converging the ground-state and computing a
sufficient number of virtual orbitals, we computed the singly-excited
(i.e., single particle-hole excitations) many-body excitations by
Casida's method. Formally this procedure amounts to approximating the
excited many-body states $|\Phi_\alpha \rangle$ by
\begin{equation}
  \label{eq:mbstates}
  |\Phi_\alpha  \rangle = \sum_{i \in \mathrm{occ}} \sum_{j \in \mathrm{virt}} A^{\alpha}_{i j}
  \hc^\dagger_{j} \hc_i |\Phi_0 \rangle \ ,
\end{equation}
where $|\Phi_0 \rangle$ denotes the determinant built by the
ground-state KS orbitals and $\hc_i$ ($\hc^\dagger_i$) is the
annihilation (creation) operator with respect to the KS basis. The
particle-hole amplitudes $A^{\alpha}_{i,j}$ are determined by Casida's
equation based on linear response. The major approximation hereby is
related to the exchange-correlation (xc) kernel $f_\mathrm{xc}(\vec
r,\vec r^\prime;\omega)$, defined as the functional derivative of the
KS potential with respect to the density. We use the
local-density approximation (LDA) for the xc kernel, as it is local
and (within adiabatic TDDFT) frequency-independent. Casida's
equation is thus transformed into an eigenvalue problem. We computed
the Casida vectors $A^{\alpha}_{i j}$ with {\sc octopus}, taking 75
occupied and 60 virtual orbitals into account, yielding well-converged
results for excitation energies up to 10\,eV. For testing purposes we
also computed the binding energies of the virtual orbitals analogously
to Ref.\,\cite{mignolet_probing_2013}. We obtained very similar
results for the low-lying states relevant for the present experiments.

\begin{figure}[t]
  \begin{center}
  \includegraphics[width=0.6\columnwidth]{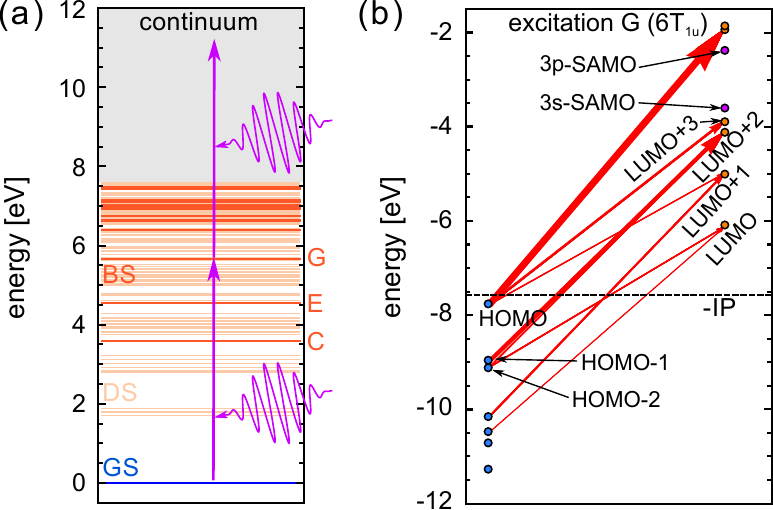}
  \end{center}
  \caption{(a) Excitation spectrum of the C$_{60}$ molecule starting
    from the ground state (GS). Optically accessible excited states
    are denoted as bright states (BS), while excitations with
    vanishing dipole transition moment are referred to as dark states
    (DS). The two-photon REMPI experiment is sensitive to BSs between
    5 and 6\,eV. The first three excitations are labeled according to
    ref.\,\cite{smith_comparison_1996}. (b) Excitation G resolved with
    respect to the constituting occupied (blue), virtual (orange) and
    SAMO (purple) electronic states.  The thickness of the red arrows
    is proportional to the corresponding weight
    $|A^{\alpha =\mathrm{G}}_{i j}|^2$.
    \label{fig1}}
\end{figure}

The energies of the obtained many-body excitations are shown in
Fig.\,\ref{fig1}(a), where we distinguish the states with vanishing
dipole transition moment (these we refer to as dark states, DS) from the
ground state GS, and optically accessible states (bright states, BS).
The onset of the visible to UV (UV-vis) optical absorption is well
documented (e.g., cf.\ Ref.\,\cite{smith_comparison_1996} for a
review and a comparison with previous experiments). Three distinct
absorption peaks (labeled according to literature as C, E and G bands,
respectively) are found in the UV-vis region. An overall agreement
between our calculations of the optical absorption with the
experimental results is found, though the excitation energies are
slightly underestimated (which is typical for DFT calculations).
Note, the specific experimental conditions may affect the spectral
positions of the absorption peaks (as discussed in
Ref.\,\cite{smith_comparison_1996}). In particular, the present
experiment probes the optical properties of isolated molecules by
ultrafast pulses and thus potentially eliminates energy-loss channels
(e.g.\ collisions and inter- or intra molecular decay) that might
shift the absorption peaks to higher energy. For these reasons, we
base the subsequent calculations on our DFT results without any
adjustments. As detailed below, very good agreement to the present
experiment is achieved, justifying this approach \emph{a posteriori}.

\begin{figure}[t]
  \begin{center}
  \includegraphics[width=0.9\columnwidth]{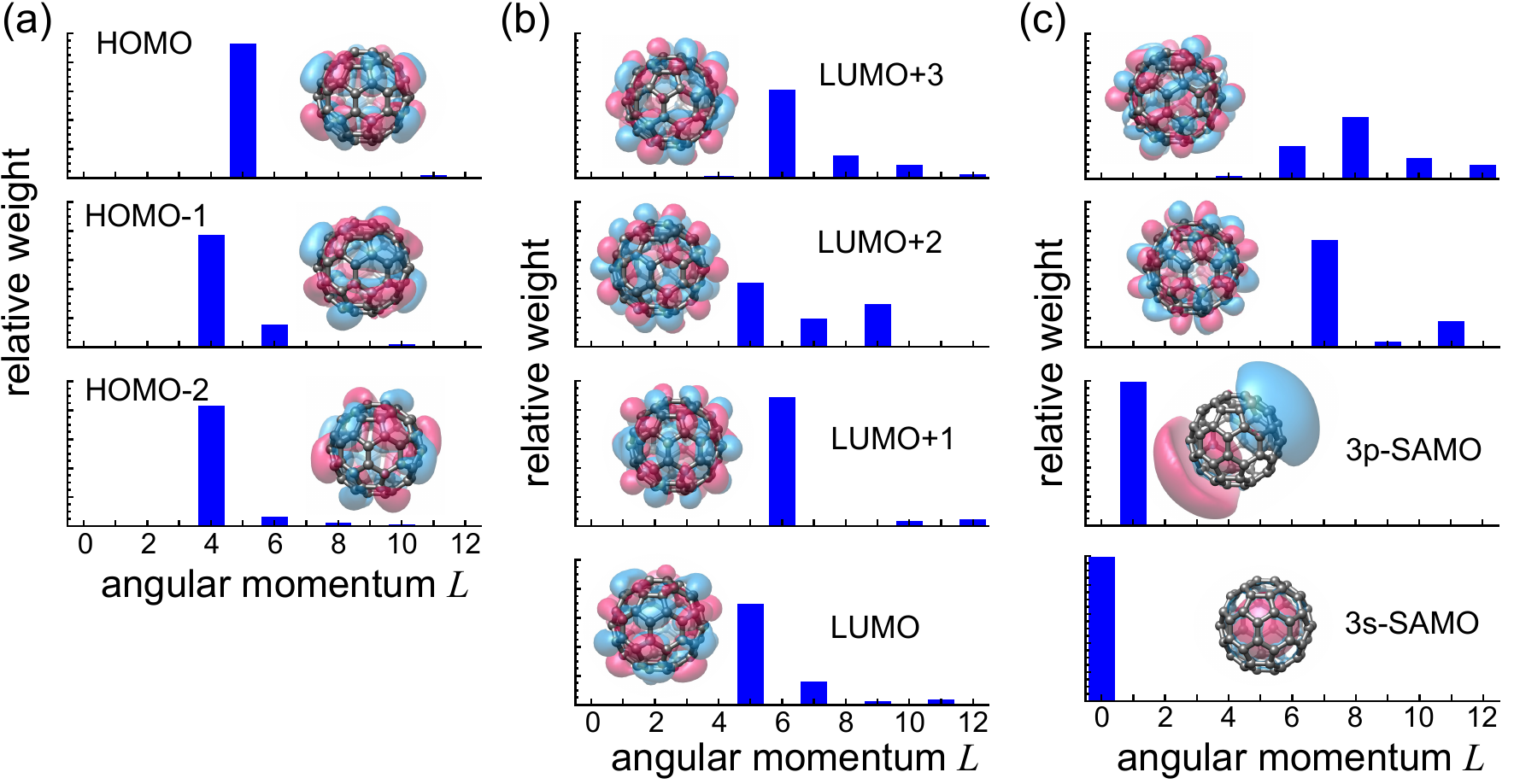}
  \end{center}
  \caption{Occupied and virtual KS states close the negative
    ionization potential (IP) ordered (from bottom to top) according
    to Fig.\,1(b).  The respective orbital character is illustrated by
    the relative weight of each angular-momentum component $L$. Only
    one representative of degenerate states is shown. (a) Occupied
    states, (b) virtual states below the SAMOs, and (c) orbitals
    including the SAMOs and higher states.}
    \label{fig2}
\end{figure}

According to our calculations, excitation G represents one triply
degenerate many-body state, which is identified as 6T$_{1u}$
excitation\,\cite{bauernschmitt_experiment_1998}. The projection onto
the KS basis is depicted in Fig.\,\ref{fig1}(b), where we illustrate
the relative weight of the excitation from occupied ($i$) to virtual
($j$) orbitals by the thickness of the corresponding arrows. The
relevant orbitals are also presented in Fig.\,\ref{fig2}. As becomes
clear from Fig.\,\ref{fig1}(b) and Fig.\,\ref{fig2}, excitation G is
predominantly composed of the transitions from (i) HOMO to virtual
states above the first SAMOs with dominant angular momentum of $L=8$
and $L=6$, and (ii) HOMO-1 to LUMO+2. The angular-momentum analysis of
the individual orbitals (Fig.\,\ref{fig2}) clarifies why the transition
to the many-body state associated to excitation G is optically
allowed. Analogously one can conclude that neither the 3s--SAMO nor
the 3p--SAMO can be populated in a direct single-photon dipole
transition from the HOMO.  This is consistent with a full symmetry
analysis of the initial orbitals and the \emph{ab initio}
calculations.

\section{Experimental setup}\label{sec:exp}
\begin{figure}[tb]
  \begin{center}
  \includegraphics[width=0.7\columnwidth]{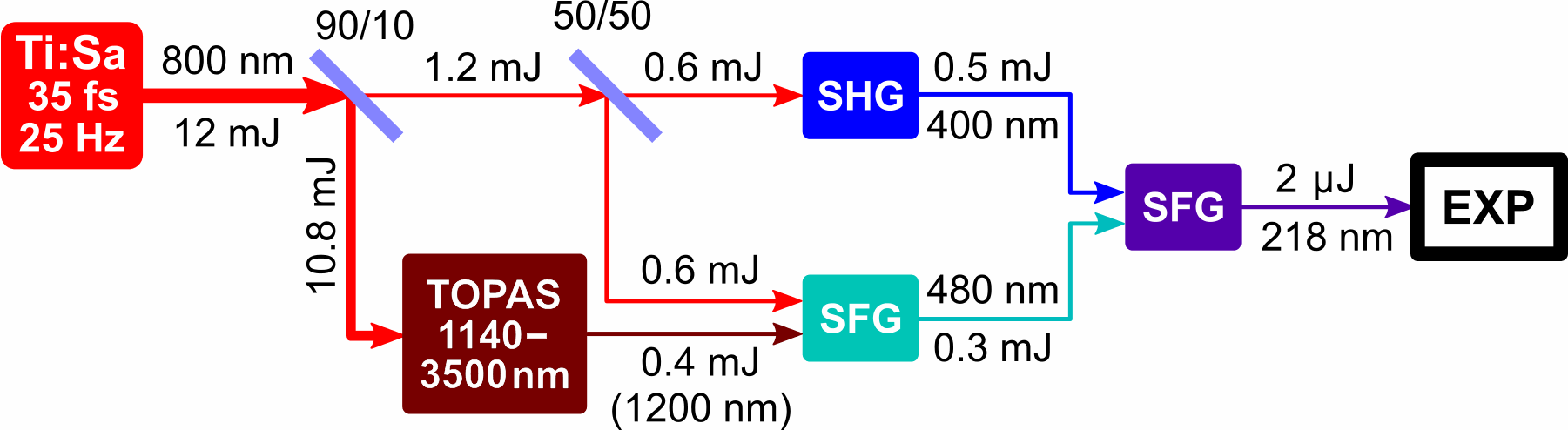}
  \end{center}
  \caption{A schematic layout of the femtosecond laser pulse
    generation scheme in the ultraviolet spectral range.}
  \label{fig4}
\end{figure}

In order to study electronic transitions into highly excited many-body
states close to the ionization potential, fs pulses at ultraviolet
(UV) frequencies are required\,\cite{johansson_visible_2013}.  The
present time-resolved mass-spectrometric study is based on
second-order UV autocorrelation making use of a time-of-flight (TOF)
mass spectrometer and state-of-the-art nonlinear optics. The laser
setup for generating fs pulses in the range of 216--222\,nm comprises
a Ti:Sa laser system, an optical parametric amplifier (OPA) and
several frequency mixing stages. The outline of the system is sketched
in Fig.\,\ref{fig4}. The Ti:Sa laser (Amplitude Technologies) is the
backbone of the overall generation scheme and provides 35\,fs (FWHM)
pulses with a pulse energy of 12\,mJ behind the compressor at a
repetition rate of 25\,Hz and 800\,nm central wavelength. This output
is split into several arms. First, the beam is split in a 90:10
ratio. The more intense fraction is sent to a commercial OPA (Light
Conversion, TOPAS-C + HE-TOPAS). The OPA is continuously tunable in
the infrared spectral range (1140--3500\,nm) and pumped by the Ti:Sa
laser. For the subsequent frequency conversion in the UV the output of
the OPA is tuned to 1200\,nm. The 11\,mJ of the 800\,nm pump pulse are
converted to ${\approx}\,0.4$\,mJ at this wavelength. The low-intensity
fraction of the Ti:Sa laser (10$\%$) is equally split into two
branches. One half is recombined with the 1200\,nm output of the TOPAS
in a $\upbeta$-barium borate (BBO, 0.2\,mm thick) crystal to generate
480\,nm light by sum frequency generation (SFG). The second half is
frequency doubled in another BBO, and then together with the 480\,nm
beam directed to the third SFG stage (40\,$\upmu$m thick) to finally
generate the 5.7\,eV (218\,nm) photons. The UV pulse energy was
measured using a calibrated XUV photodiode and a pyro detector to be ${\approx}\,2$\,$\upmu$J.

The output UV beam is directed into a vacuum chamber where it is split
into two pulses by a reflective split-and-delay unit (SDU) in order to
generate two synchronized pulse replicas. The complete nonlinear
optical setup was simulated using the software package LAB2\,\cite{schmidt_lab2-virtual_2016} including dispersion induced by UV pulse
propagation in air and through the 2\,mm thick entrance window of the
vacuum chamber. According to the calculation the 218\,nm pulse
duration in the interaction region is of the order of 100\,fs FWHM
with a spectral bandwidth of 2.8\,nm. A coarse cross-correlation
measurement performed between the 400\,nm and 480\,nm pulses of
150\,fs FWHM supports the derived UV laser beam parameters.

The SDU consists of a Si split-mirror with one half mounted on a delay
stage which can displace the mirror along its normal to set the time
delay between the pump and probe pulses. The SDU is followed by a
focussing mirror which spatially overlaps the two pulse replicas in
the laser--sample interaction area. The laser beams are focused onto
the C$_{60}$ molecular beam with a spherical mirror ($f=300$\,mm). Its
reflectivity is above 80\% in the 200--245\,nm range. The beam waist
in the interaction area is on the order of 150\,$\mathrm{\mu m}$. The
maximum peak intensity reached in the experiments is approximately
$3.5\times10^{10}$\,W/cm$^2$, which is derived from first principles
based on Gaussian beam propagation, pulse energy, pulse duration and
the far-field laser profile.

The molecular beam is produced by evaporation of gold grade C$_{60}$
powder in a resistively heated oven at 775\,K. The UV laser beams are
focused perpendicular to both, the effusive molecular beam and the TOF
spectrometer axis. The ions created in the intersection volume are
extracted by a static electric field (Wiley--McLaren configuration\,\cite{wiley_timeflight_1955}), directed onto multichannel plates, and
finally counted after amplification and discrimination by a digital
oscilloscope. The mass resolution of the TOF spectrometer is 0.2\% at
M/q=720.

\section{Two-photon photoemission}
\subsection{Excitation energy dependence \label{sec:2ppe}}

\begin{figure}[t]
  \begin{center}
  \includegraphics[width=0.6\columnwidth]{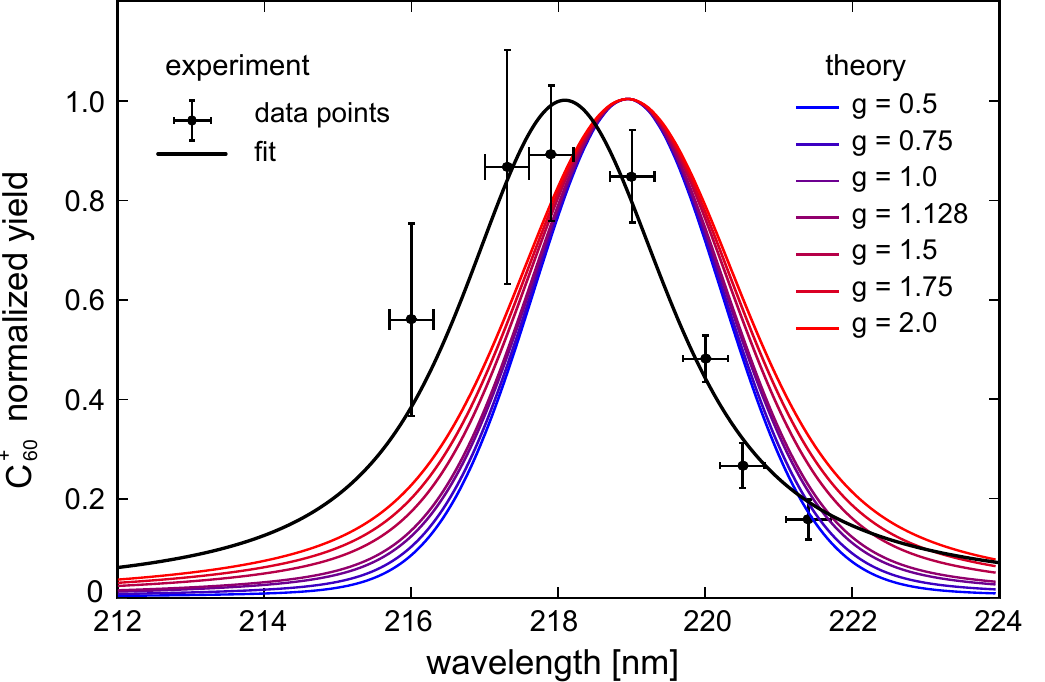}
  \end{center}
  \caption{Normalized C$_{60}^+$ ion yield (black scatter) as a
    function of excitation wavelength showing the resonance-enhanced
    two-photon photoionization. The fit to the data points (black
    line) is a convolution of a Gaussian profile representing the
    laser pulse spectrum and a Lorentzian profile,
    representing the natural linewidth of the resonance,
    respectively. The experimental data is compared with calculations
    (see Subsec.\,\ref{subsec:theovsexp}) for different values of the
    electron-vibron coupling strength $g$.}
\label{fig5}
\end{figure}

The photoionization signal recorded for resonant excitation is
enhanced compared to an experiment performed off-resonance. The
spectral width of the resonance yields information on the excited
state lifetime. A UV wavelength scan was performed by tuning the IR
wavelength of the OPA. The populated many-body state in the neutral
molecule is subsequently ionized during the pulse duration of
100\,fs. A full mass spectrum is accumulated over ${\approx}\,250$ laser
shots for each excitation wavelength. The C$_{60}^+$ yield was
normalized to the relative pulse energy monitored by a photodiode. The
C$_{60}^{+}$ ion yield as a function of laser wavelength in the range
of 216--222\,nm is shown in Fig.\,\ref{fig5}. The cut-off at 216\,nm
corresponds to the OPA's lower wavelength limit of
1140\,nm. Relatively large error bars at short wavelengths result from
the corresponding low pulse energy and thus poor statistics. The
C$_{60}^+$ signal disappears for excitation wavelengths longer than
222\,nm, thus representing the low-energy threshold of the
resonance. The wavelength scan clearly indicates resonance-enhanced
two-photon photoionization at
$\lambda_\mathrm{exc} = 218{\pm}0.5$\,nm. The observed width of the
REMPI signal is of the order of 3.65\,nm (94\,meV) pointing towards
ultrafast ionization within a lifetime that can be as short as
$10^{+5}_{-3}$\,fs. The lifetime estimate is derived from the deconvolution of the observed resonance with a Gaussian laser pulse spectrum of 2.8\,nm (FWHM) and a Lorentzian describing the homogeneous broadening.

\subsection{Pump--probe delay dependence}\label{sec:relax}
\begin{figure}[tb]
  \begin{center}
	\includegraphics[width=0.6\columnwidth]{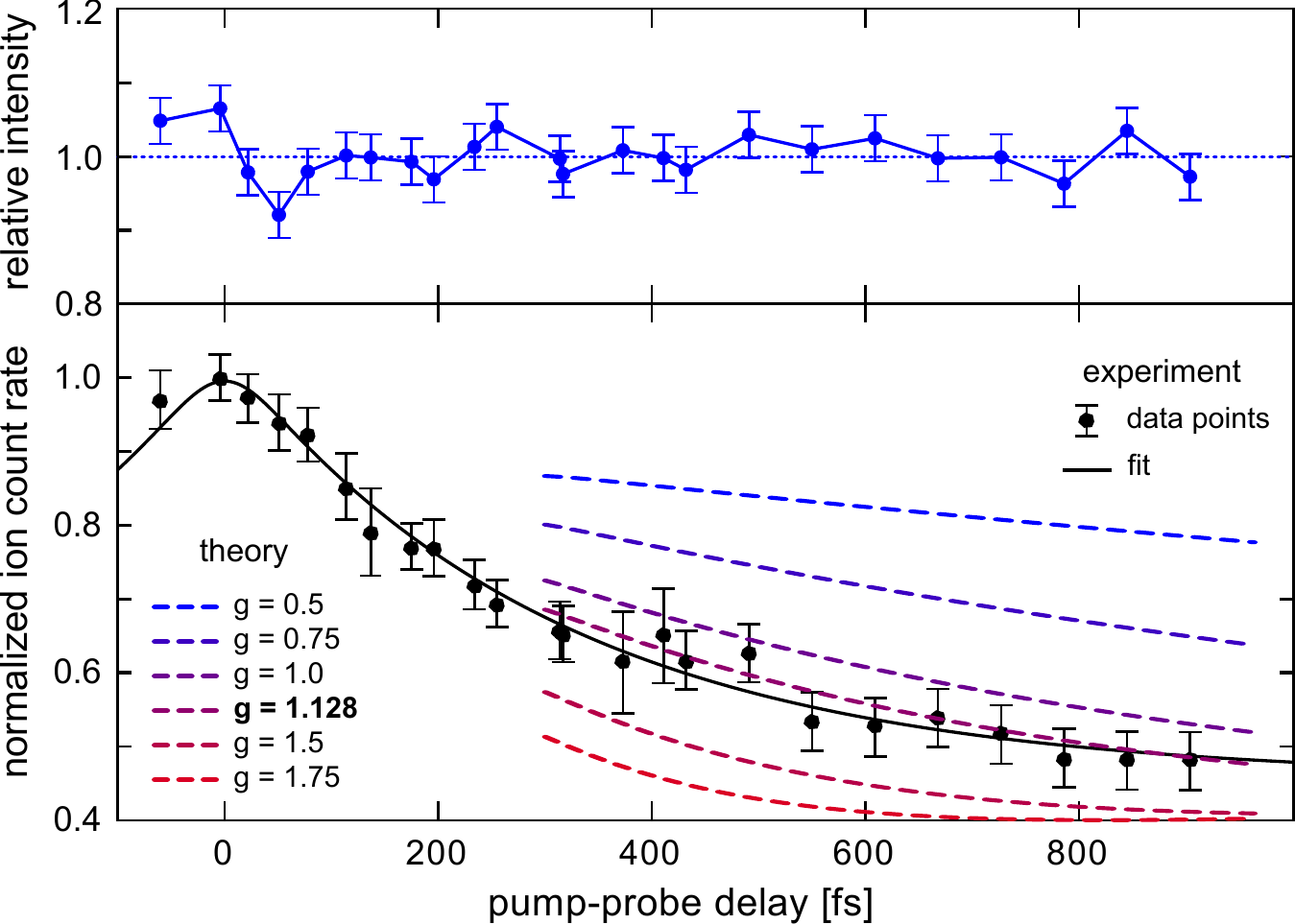}
  \end{center}
  \caption{Top graph: normalized stray light signal on the TOF
    detector induced by the excitation pulse as a function of the
    pump-probe delay, which monitors the UV laser stability throughout
    the experiment. The straight dotted line designates 1. Bottom
    graph: normalized C$_{60}^+$ counts as a function of pump-probe
    delay for resonant excitation at $\lambda_\mathrm{exc} =
    218$\,nm. The black scatter are the data points and the curve is a
    fit obtained from eq.\,\eqref{eq:fitFun} (for details see the
    text). The normalized ion counts derived from theory are compared
    to the experimental results for different scaling factors $g$ of
    the electron-vibron interaction strength (see
    Subsec.\,\ref{subsec:theovsexp}) in the limit of long pump-probe
    delays ($\geq 300$\,fs). The bold value indicates the best
    agreement between theory and experiment.}
	\label{fig8}
\end{figure}

Time-resolved mass spectrometry traces the excited state dynamics in
neutral C$_{60}$ molecules directly in the time domain. The transient
electronic structure is initiated by a UV 100\,fs pump pulse and
followed (probed) by a delayed pulse replica that photoionizes the
molecule (see Fig.\,\ref{fig1}(a)). The single-shot mass spectra are
taken at varying delay times between the UV pulses ranging from
-60\,fs to 900\,fs with ${\approx}\,3000$ laser shots per delay
point. The time-dependent on-resonance ion signal shown in
Fig.\,\ref{fig8} is derived by taking the average number of
C$_{60}^{+}$ counts for each delay and normalizing it to the relative
pulse energy monitored as the UV stray light peak in the TOF
spectrum. The pump--probe scan is repeated two times.

In the most general case the ion signal from a three-level system with
a transient intermediate electronic state exposed to resonant
excitation will result from two excitation pathways: direct two-photon
photoionization from the ground state to the continuum and the
ionization via the transient state (REMPI). Therefore, the total ion
signal $S_\mathrm{tot}$ can be described as a sum of three components\,\cite{knoesel_ultrafast_1998}:
the coherent term $S_\mathrm{ac}$ (coherent artifact), the
incoherent term $S_\mathrm{inc}$ and a constant background $a_\mathrm{bg}$. The
coherent artifact reflects the direct nonlinear ionization process and
is proportional to the autocorrelation (AC) function of the laser
pulse:
\begin{equation}
S_\mathrm{ac}(\tau) = \int\limits_{-\infty}^{+\infty} I(t-\tau)I(t) dt
\end{equation}
where $I(t)$ is the laser intensity and $\tau$ is the delay between
the pump and probe pulses. The incoherent term $S_\mathrm{inc}$
carries information about the population dynamics of the transient
state. It is a convolution of the laser pulse AC with a symmetric
decay function. In case of an exponential decay with a characteristic
time constant $\tau_\mathrm{vib}$ the incoherent term is given by:
\begin{equation}
S_\mathrm{inc}(\tau) = \int\limits_{-\infty}^{+\infty} S_\mathrm{ac}(t-\tau)\exp(-|t|/\tau_\mathrm{vib}) dt
\end{equation}
The constant background $a_\mathrm{bg}$ consists of contributions from
each excitation pulse individually and is independent of the
pump--probe delay. The total signal reads as:
\begin{equation}
S_\mathrm{tot}(\tau)=a_\mathrm{ac}S_\mathrm{ac}(\tau)+a_\mathrm{inc}S_\mathrm{inc}(\tau)+a_\mathrm{bg}
\label{eq:fitFun}
\end{equation}
with $a_i$ being the relative amplitudes of the different
components. In general the amplitudes have a ratio depending on the
spatial overlap of the two pulses and the ionization pathway of the
system. To extract $\tau_\mathrm{vib}$ from the measurement, the
experimental data is fit by the least squares method using expression
\eqref{eq:fitFun}. The background is set to $a_\mathrm{bg}=1$ and the
laser pulses are assumed to be Gaussian with
$\tau_\mathrm{FWHM}=100$\,fs. Other variables, i.e., $a_\mathrm{ac}$,
$a_\mathrm{inc}$ and $\tau_\mathrm{vib}$, are free fit parameters. The
best fit curve (black line in the bottom graph of Fig.\,\ref{fig8})
yields a time constant \mbox{$\tau_\mathrm{vib} = 400\pm100$\,fs} (95\%
confidence band) and an amplitude ratio \mbox{$a_\mathrm{ac}:a_\mathrm{inc}:a_\mathrm{bg}=0.24:1.13:1$}. The laser
intensities in the interaction region in the present experiment are as
low as $3.5\times10^{10}$\,W/cm$^2$ which makes the contribution from
the direct (nonlinear) two-photon process small. The ratio
$a_\mathrm{inc}:a_\mathrm{bg}$ is close to 1 as expected for single
photon ionization from an occupied transient state.

The observed exponential time constant
$\tau_\mathrm{vib} = 400\pm100$\,fs is significantly longer than the
characteristic electron--electron interaction time derived from
pump--probe spectroscopy\,\cite{shchatsinin_c60_2006} and pulse
duration dependent studies\,\cite{campbell_above_2000,hansen_thermal_2003} in the optical spectral
range. It seems that electron thermalization mediated by inelastic
electron--electron collisions on a time scale of $50{-}100$\,fs does
not play a key role when high lying correlated many-body states are
excited directly at rather low peak intensity. In the following a
detailed theoretical description of the observed resonance and its
time-dependent structure is discussed.

\section{Simulations\label{sec:T2ppe}}

In order to the describe the response of the molecule upon pulsed
laser irradiation, the interplay between the electronic excitations
and the vibrationally hot molecule has to be taken into account (for
an overview on this topic we refer to the book\,\cite{chancey_jahn-teller_1997} and further references therein). In
full generality, an \emph{ab initio} description for both the
electronic and nuclear degrees of freedom and their coupling is not
feasible currently. Hence, one has to rely on suitable
approximations. In an Ehrenfest approach the electrons are described
on the level of TDDFT and the nuclei are subject to classical
equations of motion due to the forces exerted by the electron
distribution. Despite the success of this molecular dynamics approach\,\cite{leszczynski_handbook_2012} for predicting vibrationally-assisted
charge-transfer processes\,\cite{pittalis_charge_2015} in photovaltic
molecules, the Born-Oppenheimer (BO) approximation is an inherent
limitation hereby. For excited-state properties, where the nuclear and
electronic dynamics are strongly mixed, BO-type molecular dynamics is
not predictive. Molecular dynamics beyond the BO approximation\,\cite{ben-nun_ab_2000,nyman_quantum_2000,schlegel_exploring_2003} has
been employed for the C$_{60}$ molecule\,\cite{fischer_orientation_2013}; merging such schemes with a treatment
of the electronic excitations beyond the KS level is computationally
too demanding for our system.
Alternatively, one can treat the electrons in a single-particle atomic basis within a
tight-binding model\,\cite{zhang_laser-induced_2003}, removing the adiabaticity
constraint with respect to the many-body states. Besides
the inevitable empirical ingredient, such theory is also not directly
compatible with the \emph{ab initio} description of the many-body
states in Sec.\,\ref{sec:optabs}, i.\,e. electronic correlations can
only be taken into account by great effort.

\subsection{Initial laser-induced dynamics}\label{sec:dyn}

\begin{figure}[t]
  \begin{center}
  \includegraphics[width=0.6\columnwidth]{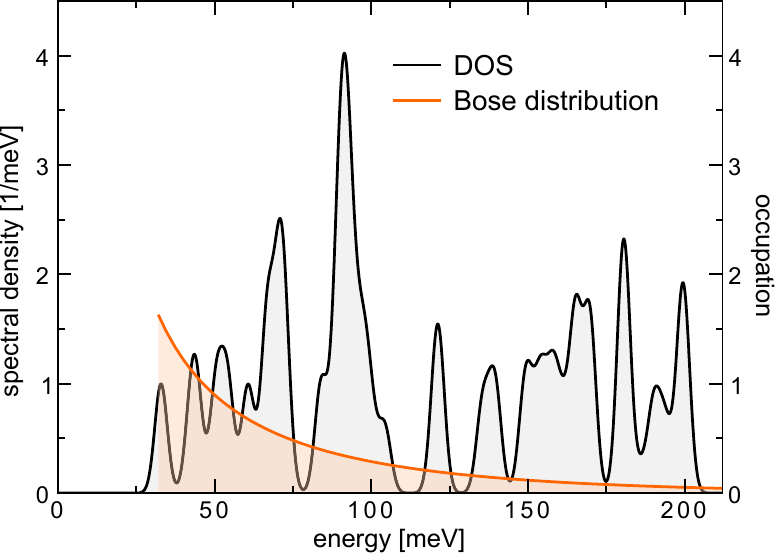}
  \end{center}
  \caption{Vibrational density of states (DOS) along with the
    occupation according to the Bose distribution for $T=775$\,K.}
    \label{fig3}
\end{figure}

In order to elucidate the laser- and vibration-induced dynamics we
take a different angle. Since a considerable amount of energy is
stored in thermally activated vibrations, which can only be
transferred to the electronic subsystem in small portions, the vibrons
can be treated as an effective heat bath for the electrons.
A similar model has successfully been employed for incorporating the
influence of the vibrations on charge-transfer processes in organic
photovoltaic systems based on C$_{60}$\,\cite{tamura_quantum_2012}.
To construct an appropriate model for our case, several ingredients
are required. For the vibrations we restrict ourselves to the harmonic
approximation of the bottom of the BO surfaces. The vibronic
eigenmodes along with their eigenfrequencies and reduced masses were
computed using the {\sc Octopus} code, as well.  The resulting density
of states (DOS) of the vibronic modes is shown in Fig.\,\ref{fig3}. Our
results compare very well with those tabulated in the literature, for
instance table 6.2 in the book\,\cite{chancey_jahn-teller_1997}.

As inferred from Fig.\,\ref{fig3} the high-energy modes (which affect
the electrons the most) are only weakly populated for
$T=775$\,K. Therefore, the oscillations of the nuclei around their
equilibrium positions can be considered small. Hence, the
Herzberg-Teller (HT) expansion\,\cite{koppel_jahn-teller_2009} of the
full Hamiltonian, including electrons and nuclei, yields a reasonable
description for both subsystems and their interaction. The first-order
HT Hamiltonian amounts to approximating the electron--vibron coupling
as linear in the mode amplitudes $\hat{Q}_\nu$. On the KS level, the
electron--vibron matrix elements are thus given by
\begin{equation}
  \label{eq:elvibmatks}
  k^\nu_{ij}=\langle \phi_i | \frac{\partial v^{\mathrm{KS}}}{\partial Q_\nu} |
  \phi_j\rangle \Big|_{Q_\nu = 0} \ .
\end{equation}
Details on the evaluation of eq.\,\eqref{eq:elvibmatks} are provided in
\ref{app:elvib}. Since we are opting for a model in the many-body
basis of the excitations discussed in Sec.\,\ref{sec:optabs}, the
matrix elements \eqref{eq:elvibmatks} are transformed according to
eq.\,\eqref{eq:mbstates} (see \ref{app:elvib}). We thus obtain
the model Hamiltonian (atomic units are used throughout)
\begin{equation}
  \label{eq:Hmodel}
  \hat{H}(t) = \hat{H}_\mathrm{el}(t) + \hat{H}_\mathrm{el-vib} + \hat{H}_\mathrm{vib} \ ,
\end{equation}
where
\begin{equationarray}{rcl}
  \label{eq:hel}
  \hat{H}_\mathrm{el}(t)&=&\sum_\alpha E_\alpha |\Phi_\alpha \rangle \langle \Phi_\alpha |+
 f(t) \sum_{\alpha \beta} M_{\alpha \beta} |\Phi_\alpha \rangle \langle \Phi_\beta |\ , \\
   \label{eq:helvib}
  \hat{H}_\mathrm{el-vib}&=& g \sum_{\alpha\beta} \sum_\nu K^\nu_{\alpha \beta}
  |\Phi_\alpha \rangle \langle \Phi_\beta | \hat{Q}_\nu \ , \\
   \label{eq:hvib}
  \hat{H}_\mathrm{vib}&=&\frac12\sum_\nu\left(\frac{\hat{P}^2_\nu}{M_\nu}
    + k_\nu \hat{Q}^2_\nu \right) \ .
\end{equationarray}
Here, $M_{\alpha \beta}$ are the dipole transition matrix elements
from our TDDFT calculations, while $f(t)$ comprises the time-dependent
fields. The prefactor $g$ in eq.\,\eqref{eq:helvib} is introduced as an
overall scaling factor for the strength of the vibronic
coupling. Ideally, $g=1$ should be fixed; however, due to the
perturbative description resulting from the HT expansion,
the electron-vibron interaction might be underestimated. Hence, $g$ is
kept as a parameter.

Instead of describing the dynamics of the full density matrix
according to Hamiltonian\,\eqref{eq:Hmodel} (which is a formidable
task), we treat the vibrations, as explained above, as a heat
bath. This allows to obtain a master equation for the density matrix
in the electronic subspace only. Here we assume Markovian dynamics and
thus employ the Lindblad master equation, following the standard
derivation and formulation from Ref.\,\cite{breuer_theory_2002}. More
details are presented in \ref{app:lindblad}. This procedure requires
an additional parameter: the vibronic broadening
$\eta$, which corresponds to the lifetime of the vibrational modes.
Note that the laser-induced dynamics is, besides the electron-vibron
interaction, basically treated on \emph{ab initio} level endorsing the
predictive power of the approach. Note that the electron-vibron
coupling\,\eqref{eq:helvib} includes, in principle, Jahn-Teller and
Herzberg-Teller couplings, which where identified as the main
mechanisms for vibrational coupling in
fullerenes\,\cite{sassara_visible_1997,menendez_vibrational_2000,orlandi_electronic_2002}.

\begin{figure}[t]
  \centering
  \includegraphics[width=0.9\columnwidth]{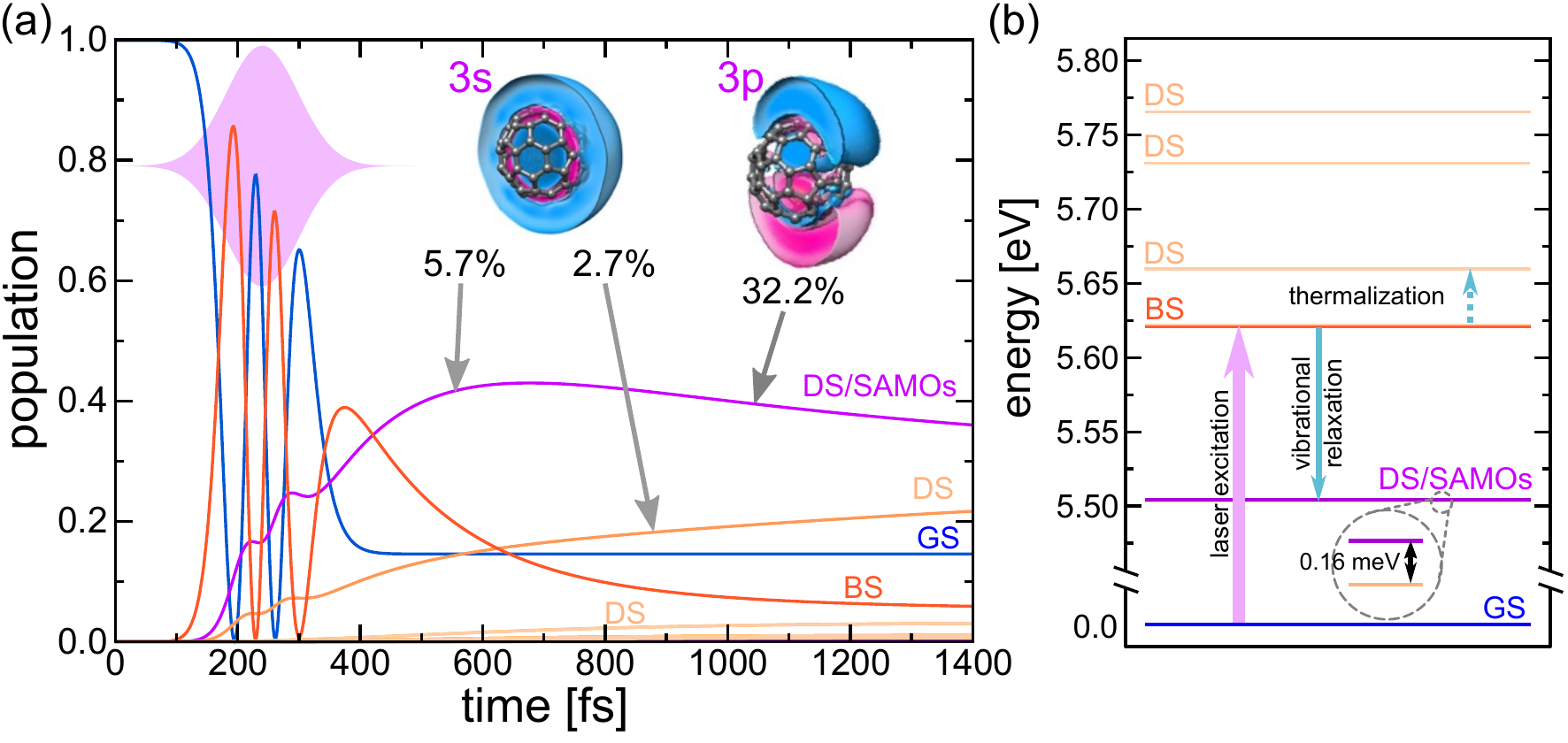}
  \caption{(a) Population dynamics induced by the pump pulse (sketched
    in the background). The color coding of ground state (GS), bright
    states (BS) and dark states (DS) is identical to
    Fig.\,\ref{fig1}(a). The insets show the weight of the 3s-- and
    3p--SAMOs in the dominantly populated states. The DSs involving
    the strongest SAMO excitations is highlighted by the purple
    color. (b) Dominant population mechanisms for the dynamics in
    (a).} \label{fig6}
\end{figure}

The time evolution of the occupation of the states depicted in the
level scheme Fig.\,\ref{fig1}(a) is presented in Fig.\,\ref{fig6}(a). The driving pulse $f(t)$ is chosen as a Gaussian pulse with a
FWHM of 100\,fs as in the experiment, while the central frequency is
adjusted to the vertical excitation energy $\Delta E = 5.66$\,eV
between ground state and the BSs corresponding to the G peak. The peak
amplitude amounts to the intensity of $3.5\times
10^{10}$\,W/cm$^2$. The values for $g$ and $\eta$ are chosen to match
the time-dependent pump-probe signal observed in the experiment (see
Subsec.\,\ref{subsec:theovsexp}).

As one can infer from Fig.\,\ref{fig6}, the
population transfer between the ground state and the bright excited
states is clearly not in a perturbative regime, as two Rabi cycles are
apparent during the 100\,fs UV pulse interaction. The relaxation
dynamics, transferring part of the excitation to the dark states,
takes place on two time scales: for short delays one can observe a
rapid energy transfer, while for longer times the distribution
thermalizes. The depletion dynamics of the laser-excited BS primarily
takes place due to the coupling to two lower-energy states at
$\approx\,5.5$\,eV (see Fig.\,\ref{fig6}(b)). Closer inspection reveals
that these DSs involve the excitation of the 3s and the 3p SAMOs; the
respective weights are given in Fig.\,\ref{fig6}(a). This relaxation mechanism is
the dominant consequence of the electron-vibron coupling. This
behavior is expected, as dissipation pathways are generally preferred
as compared to bath-induced excitations. These thermalization
processes are hence less pronounced and occur on a longer time scale.

\subsection{Pump-probe dynamics}

In order to compute the pump-probe signals from the dynamics of the
density matrix $\rho_{\alpha \beta}(t)$, as discussed above, an
extension to the scattering states is required. However, a
straightforward implementation of the Lindblad equation including both
bound and unbound many-body states is not feasible. This is due to the
large dimension of the Hamiltonian after the inevitable discretization
of the continuum. Hence, we opt for a perturbation description which
allows to compute the pump-probe dynamics from the time-dependent
density matrix without incorporating the ionization dynamics
explicitly. This is achieved by a method known from time-resolved
photoelectron spectroscopy\,\cite{kemper_effect_2014}. Adopting a
straightforward derivation for the many-body case, one obtains
\begin{equation}
  \label{eq:tdspec1}
\hspace*{-2cm}  N_{\vec k} \propto \mathrm{Re}\sum_{\alpha \beta} \int^\infty_{-\infty}\!\dd t\!\int^t_{-\infty}\!\dd t^\prime \,
  F_\mathrm{probe}(t)F_\mathrm{probe}(t^\prime)e^{-\iu(E_\alpha + \omega - E^+_\beta -\epsilon_{\vec k})(t-t^\prime)}
  |M_{\beta \vec{k},\alpha}|^2 \rho_{\alpha \alpha}(t^\prime)
\end{equation}
for the number $N_{\vec k}$ of released photoelectrons with momentum
$\vec{k}$ and energy $\epsilon_{\vec k}$. The probe laser pulse is
assumed as $f_\mathrm{probe}(t)=F_\mathrm{probe}(t)e^{-\iu \omega t}$
with the pulse envelop $F_\mathrm{probe}$. The matrix element
$M_{\beta \vec k,\alpha}=\langle \Phi_{\beta,\vec
  k}|\hat{D}|\Phi_\alpha \rangle$ ($\hat{D}$ denotes the dipole
operator) describes the transition from the intermediate states
$|\Phi_\alpha \rangle$ to the final states
$|\Phi_{\beta,\vec k}\rangle$ with one photoelectron and the ion state
labeled by $\beta$ (energy $E^+_\beta + \epsilon_{\vec k}$). Note
that only the population of excited intermediates states and not the full density matrix enters
eq.\,\eqref{eq:tdspec1}. This is an approximation, which relies on the
fact that pathway interferences play only a minor role for the
considered two-step ionization process.

For the excited state with one electron in the continuum state $|\vec k \rangle$,
$|\Phi_{\beta, \vec{k}} \rangle$, we write the usual anti-symmetrized product ansatz
\begin{equation}
  |\Phi_{\beta, \vec{k}} \rangle = \hc^\dagger_{\vec k} |\Phi^+_{\beta} \rangle \ ,
\end{equation}
where $|\Phi^+_{\beta} \rangle$ is an eigenstate of the ionized
system with energy $E^+_\beta$. In this case the photoemission matrix element reduces to
\begin{equation}
  \label{eq:speme1}
  \langle \Phi_{\beta,\vec k}| \hat{D} | \Phi_\alpha \rangle = \langle \vec k | \hat{D}
  | \phi^\mathrm{D}_{\alpha \beta} \rangle \ .
\end{equation}
Here, $\phi^\mathrm{D}_{\alpha \beta} (\vec r)$ stands for the
corresponding Dyson orbital. As the sum over all excited states $\beta$ is
implied, a (computationally expensive) precise calculation of the
Dyson orbitals can be omitted by approximating them by simple hole
states, i.e., by assuming
\begin{equation}
  |\Phi^+_{\beta=(m,\alpha)} \rangle \approx \hc_{m} |\Phi_{\alpha}
  \rangle \ , \quad E^+_\beta \approx E_\alpha - \epsilon_m \ .
\end{equation}
Here, $\epsilon_m$ stands for the KS eigenvalue of orbital
$\phi_m(\vec r)$.
The matrix element \eqref{eq:speme1} can thus be evaluated in
terms of a superposition of matrix elements with respect to the KS
orbitals. The scattering states $|\vec k \rangle$ were computed with
respect to the spherically averaged KS potential\,\cite{schuler_disentangling_2015}, which is known to be an adequate
approximation for angle-integrated quantities. Asymptotic corrections
ensuring the $r^{-1}$ behavior are incorporated by smoothly
interpolating between the short-range and long-range regimes. Further
orthogonalization with respect to the bound KS orbitals is performed.

To reflect the experimental situation, the integration over all
photoelectron states has to be performed. Furthermore, we note that
eq.~\,\eqref{eq:tdspec1} balances spectral resolution vs. temporal
resolution in terms of the convolution of the phase factor
$e^{-\iu \Delta E t}$,
$\Delta E = E_\alpha + \omega - E^+_\beta -\epsilon_{\vec k}$, with
the envelop function. Due to very long pulses as compared to one
oscillation period, this convolution practically yields a Dirac
$\delta$-function with respect to the energy balance. Taking this into
account, the total ionization pump-probe signal
$S(\tau) = \int\!\dd \vec{k}\, N_{\vec k}$ simplifies to
\begin{equation}
  \label{eq:tdspec2}
  S(\tau) \propto \sum_{\alpha \beta} \int^\infty_{-\infty}\!\dd t\!\int^t_{-\infty}\!\dd t^\prime \,
  F_\mathrm{probe}(t-\tau)F_\mathrm{probe}(t^\prime-\tau)P_{\alpha \beta}(E_\alpha + \omega)
  \rho_{\alpha \alpha}(t^\prime) \ ,
\end{equation}
where
$P_{\alpha \beta}(\epsilon) = k\int\!\dd{\Omega}_{\vec k}
|M_{\beta \vec{k},\alpha}|^2$ with $k=\sqrt{2\epsilon}$ is
proportional to the energy-resolved ionization probability with
respect to the initial state $|\Phi_\alpha \rangle$ and final state
$|\Phi^+_\beta\rangle$. As in Sec.\,\ref{sec:relax}, $\tau$ denotes the
pump-probe delay.

\subsection{Theory vs. experiment\label{subsec:theovsexp}}

The pump-probe signal based on the laser-driven and vibron-coupled dynamics of the density matrix can now
be compared to the experiment. We remark that Eq.\,\eqref{eq:tdspec2}
assumes that the two pulses can be separated and does not account for
the scenario of overlapping pulses. We thus limit the comparison
between the theoretical calculations and the experiment, presented in
Fig.\,\ref{fig8}, to the region of exponential decay for
$\tau \geq 300$\,fs. Furthermore,  the dynamics presented in
Fig.\,\ref{fig6} indicates that the laser intensity exceeds the perturbative
regime. However, even with stronger pulses, the bright states around
$5.66$\,eV are the only accessible channels, as absorbing another
photon leads to immediate ionization. This will, however, only affect
the background signal. Therefore, we adjust the background
$S_\mathrm{bg} = S(\tau\rightarrow \infty)$ to the experiment. After
solving the Lindblad master equation for various values of the
parameters $g$ and $\eta$, we found the best fit for $g=1.128$ and
$\eta = 0.77$\,meV. The latter corresponds to a vibrational lifetime of
$\approx 5.4$\,ps, which is in accordance with previous
experiments\,\cite{hertel_ultrafast_2005}. The small deviation of the
scaling factor $g$ from unity underpins the predictive power of our
treatment. Note that also $g=1$ results in a decay dynamics which
closely resembles, apart from the background, the experimental data,
whereas varying $g$ to smaller or larger values clearly deviates from
the measurements.

We also calculated the ionization signal for varying photon energy $\omega$
and compared the resulting spectra in Fig.\,\ref{fig5}. To match the
laser spectral bandwidth to the experiment, the obtained curves were convolved with a Gaussian
having a FWHM of 2.8\,nm.
The theoretical spectra are centered around the vertical excitation energy, which is
lower than what is observed in the experiment. This kind of
underestimating of bandgaps and thus excitation energies is typical
for most DFT calculations and for the C$_{60}$ molecular, in
particular\,\cite{bauernschmitt_experiment_1998}. However, the
theoretical and the experimental peak differ only by $\approx\,10$\,meV.
 Generally, both the experimental as well as
the theoretical spectra are considerably sharper as compared to
optical absorption measurements. This is a major advantage of
the current experimental setup: the dipole excitation dominates over
possible loss channels, if the pulse strength is increased, which
leads to a narrow resonance.

\subsection{Steady-state vs. time-dependent picture}

The interpretation of the present data on the relaxation dynamics
requires the full picture including both, correlated electronic and
nuclear degrees of freedom and their interactions\,\cite{lezius_nonadiabatic_2001,lezius_polyatomic_2002}.
It is known that highly excited electronic and vibrational C$_{60}$ states are strongly mixed\,\cite{torralva_response_2001,zhang_laser-induced_2003}.
In turn, relaxation channels open up that depopulate the
electronically excited states by internal conversion close to conical
intersections similar to electron--phonon coupling in solid state
materials. We note that characteristic fragmentation patterns observed
in TOF mass spectra using optical lasers as a function of pump--probe
delay revealed (nonadiabatic) vibronic coupling on a time scale of
$\tau_\mathrm{vib}=200{-}300$\,fs\,\cite{shchatsinin_c60_2006},
which is in good agreement with the present findings. Furthermore,
nonadiabatic coupling of mixed valence and Rydberg states is
ubiquitous in polyatomic molecules affecting potential energy
surfaces, energy relaxation and dissociation dynamics\,\cite{mignolet_probing_2013}.
Similar processes have been considered to evoke the
loss of small neutral fragments from C$_{60}$ on a picosecond time
scale\,\cite{boyle_fragmentation_2005}.

The identification of vibronic coupling playing a key role in the
electronic energy loss of correlated many-body states may open new
vistas for optical control of charge-transport phenomena in smart
materials containing these nanospheres. For instance, coherently
induced radial symmetric ''breathing'' motion of the cage atoms
strongly impacts the structure and dynamics of the molecule\,\cite{laarmann_control_2007}.
The carbon cage and the electron system start to exchange many eV in
energy periodically on a sub-100\,fs period timescale. The coherent
oscillation prevails for several cycles\,\cite{laarmann_control_2007},
which might be interesting for novel ultrafast switching applications
in molecular electronics.

\section{Summary and Outlook}
Ultrashort pulses in the ultraviolet spectral range excite correlated
many-body states of isolated C$_{60}$ molecules below the ionization
continuum. The population is followed by subsequent UV pulses of the
same wavelength that ionize the molecules. By recording the C$_{60}^+$
ion yield as a function of time delay between pump and probe pulses we
observe an exponential decay with a time constant of $\tau_\mathrm{vib} =
400\pm100$\,fs which is explained by the coupling of electronic
excitation to nuclear motion in the neutral molecule. The initial
electronic relaxation can be as fast as $\tau_\mathrm{el}=10^{+5}_{-3}$\,fs
according to the evaluation of the resonance line width in single
pulse experiments as a function of excitation wavelength. The
experimental results are in good agreement with \emph{ab initio} calculation
of structure and dynamics including electronic correlation and
vibronic coupling. However, the UV pulse duration of 100\,fs did not
allow to observe pure electron dynamics in time-resolved
experiments. Future experimental work making use of shorter UV pulses
shall reveal the predicted laser-driven Rabi oscillation and
time-resolved transformation of the electronic orbitals, i.e., the
coupling between different electronic states.

\section*{Acknowledgment}
This research was supported by the Deutsche Forschungsgemeinschaft
through the excellence cluster ''The Hamburg Centre for Ultrafast
Imaging (CUI) -- Structure, Dynamics and Control of Matter at the
Atomic Scale'' (DFG-EXC1074), the collaborative research centres
''Light induced dynamics and control of correlated quantum systems''
(SFB 925), ''Functionality of Oxide Interfaces'' (SFB 762), the GrK~1355 and Grant Number PA 1698/1-1.

\appendix

\section{Electron-vibron matrix elements\label{app:elvib}}

In this Appendix we provide the details on how the first-order
(Herzberg-Teller) electron-vibron coupling matrix elements,
Eq.\,\eqref{eq:elvibmatks}, were computed. After the calculation of the
vibrational eigenmodes $\nu$, one readily obtains the associated
deformations of the fullerene cage. Denoting the collection of nuclear
coordinates in equilibrium by
$\{\vec{R}^{(0)}\}=(\vec{R}^{(0)}_1,\dots,\vec{R}^{(0)}_{60})$, the vibrational
distortion is characterized by
\begin{equation}
  \label{eq:displgeom}
  \{\vec{R}^{(\nu)}(q)\}
=(\vec{R}^{(\nu)}_1(q),\dots,\vec{R}^{(\nu)}_{60}(q)) \quad \mathrm{with} \quad
\vec{R}^{(\nu)}_{m}(q) = \vec{R}^{(0)}_m + q \vec{V}^{(\nu)}_m \ ,
\end{equation}
where $\vec{V}^{(\nu)}_m$ is the displacement eigenvector. Based on
Eq.\,\eqref{eq:displgeom}, we performed a DFT calculation for each
vibrational mode, fixing the magnitude of the distortions at
$\delta q=0.01$. Instead of taking derivatives of the KS potential, we
employ the equivalent formulation
\begin{equation}
  \label{eq:knuij1}
    k^{\nu}_{ij} \simeq \frac{1}{\delta q} \langle \phi_i|\hat{h}_\nu - \hat{h}_0|\phi_j \rangle \ ,\ \quad \hat{h}_\nu  = -\frac12 \nabla^{2} + v^{\mathrm{KS}}(\vec r, \{\vec{R}^{(\nu)}(\delta q)\}) \ ,
\end{equation}
where we have approximated the derivative by finite differences. The
KS Hamiltonian describing the molecule in equilibrium is denoted by
$\hat{h}_0$, while $\hat{h}_\nu$ describes the distorted molecule. For
the evaluation of Eq.\,\eqref{eq:knuij1} we insert a (approximate)
completeness relation and obtain
\begin{equation}
  \label{eq:knuij2}
  \begin{array}{l}
    k^{\nu}_{ij} \simeq \frac{1}{\delta q}\left[ \langle \phi_i|\tilde{\phi}^{(\nu)}_k \rangle \tilde{\epsilon}^{(\nu)}_k \langle \tilde{\phi}^{(\nu)}_k |\phi_j \rangle - \epsilon_i \delta_{ij} \right] \ ,\\[3mm]
    \hat{h}_\nu|\tilde{\phi}^{(\nu)}_k \rangle =\tilde{\epsilon}^{(\nu)}_k |\tilde{\phi}^{(\nu)}_k \rangle \ , \ \hat{h}_0|\phi_i\rangle=\epsilon_i |\phi_i\rangle \ .
  \end{array}
\end{equation}
Note that the overlaps
$\langle \phi_i|\tilde{\phi}^{(\nu)}_k \rangle$ incorporate the
symmetry properties of the vibronically induced transitions.

The transformation into the many-body basis is accomplished by
expressing the one-body coupling operator in second quantization,
$\hat{k}^{\nu}=\sum_{ij}k^{\nu}_{ij} \hat{c}^\dagger_i \hat{c}_j$, and
evaluating $\langle \Phi_\alpha|\hat{k}^{\nu}|\Phi_\beta \rangle$
according to the algebra of the fermionic creation and annihilation operators.

\section{Lindblad master equation \label{app:lindblad}}
\setlength{\arraycolsep}{0.2em}
For the derivation of the Lindblad master equation in the
weak-coupling limit, we follow Ref.\,\cite{breuer_theory_2002}. For the
Hamiltonian\,\eqref{eq:hel}--\eqref{eq:hvib} one obtains the following
EOM for the electronic density matrix
$\hat{\rho}(t) = \sum_{\alpha \beta} \rho_{\alpha \beta}(t)|\Phi_\alpha\rangle \langle \Phi_\beta |$:
\begin{equation}\begin{array}{rl}
  \frac{\dd}{\dd t}\hat{\rho}(t)=&-\iu \left[\hat{H}_\mathrm{el}(t),\hat{\rho}(t)\right]\\[2mm]
  &+\displaystyle\sum_{\alpha \beta \alpha^\prime \beta^\prime}\Gamma_{\alpha \beta \alpha^\prime \beta^\prime}
  \Big( \rho_{\beta \beta^\prime}(t)|\Phi_\alpha \rangle \langle \Phi_{\alpha^\prime}|
 -\frac12 \left\{ |\Phi_{\beta^\prime}\rangle \langle \Phi_{\beta}| \hat{\rho}(t) \right\}
\Big)\ .\end{array}
\end{equation}
The square (curly) brackets denote the commutator (anti-commutator). The vibronic bath enters into
\begin{equation}
  \Gamma_{\alpha \beta \alpha^\prime \beta^\prime} =\sum_\nu \gamma_\nu(E_\beta-E_\alpha)
  \delta_{E_\beta-E_\alpha,E_{\beta^\prime}-E_{\alpha^\prime}}K^\nu_{\alpha \beta}
  K^\nu_{\alpha^\prime \beta^\prime} \ ,
\end{equation}
where $\gamma_\nu(E) = (N_\mathrm{B}(E)+1)
A_\nu(E)$. $N_\mathrm{B}(E)$ denotes the Bose distribution (displayed
in Fig.\,\ref{fig3}) which accounts for the occupation of the vibronic
modes for the given temperature. The vibrational frequencies determine
the spectral function \mbox{$A_\nu(E)=2\pi \delta(E-\Omega_\nu)$}, which we
replace by the smeared form
\begin{equation}
  A_\nu(E)=\sqrt{2\pi/\eta^2}\exp[-(E-\Omega_\nu)^2/2\eta^2]
\end{equation}
to account for the finite lifetime of the vibrations.

\section*{References}
\providecommand{\newblock}{}

\end{document}